\documentclass{article}

\usepackage{arxiv}

\usepackage[utf8]{inputenc} 
\usepackage[T1]{fontenc}    
\usepackage{hyperref}       
\usepackage{url}            
\usepackage{booktabs}       
\usepackage{amsfonts}       
\usepackage{nicefrac}       
\usepackage{microtype}      
\usepackage{lipsum}		
\usepackage{graphicx}
\usepackage[square,sort,comma,numbers]{natbib}
\usepackage{doi}
\usepackage{setspace}
\usepackage[font=onehalfspacing]{caption}

\usepackage{graphicx}
\usepackage{mathptmx}
\usepackage{amsmath}
\usepackage{amssymb}

\usepackage{amsthm}
\usepackage{algorithm,algorithmic}

\newtheorem{thm}{Theorem}[section]

\newtheorem{prop}[thm]{Proposition}

\theoremstyle{definition}

\title{An invertible transform for efficient string matching in labeled digraphs}

\author{ \href{https://orcid.org/0000-0001-8145-1484}{\includegraphics[scale=0.06]{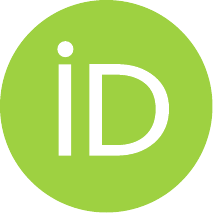}\hspace{1mm}Abhinav Nellore} \\
	Oregon Health \& Science University \\
	Portland, Oregon 97239, USA \\
	\texttt{anellore@gmail.com} \\
	\And
	\href{https://orcid.org/0000-0001-7940-4830}{\includegraphics[scale=0.06]{orcid.pdf}\hspace{1mm}Austin Nguyen} \\
	Oregon Health \& Science University\\
	Portland, Oregon 97239, USA \\
	\texttt{nguyenau@ohsu.edu} \\
	\And
	\href{https://orcid.org/0000-0003-3661-5296}{\includegraphics[scale=0.06]{orcid.pdf}\hspace{1mm}Reid F.~Thompson} \\
	Oregon Health \& Science University \\
	VA Portland Healthcare System \\
	Portland, Oregon 97239, USA \\
	\texttt{thompsre@ohsu.edu} \\
}

\renewcommand{\headeright}{}
\renewcommand{\undertitle}{A preprint}
\renewcommand{\shorttitle}{An invertible transform for efficient string matching in labeled digraphs}

\hypersetup{
pdftitle={An invertible transform for efficient string matching in labeled digraphs},
pdfsubject={cs.DS},
pdfauthor={Abhinav Nellore, Austin Nguyen, Reid F.~Thompson}
}

\begin{document}
\onehalfspacing
\maketitle

\begin{abstract}
Let $G = (V, E)$ be a digraph where each vertex is unlabeled, each edge is labeled by a character in some alphabet $\Omega$, and any two edges with both the same head and the same tail have different labels. The powerset construction gives a transform of $G$ into a weakly connected digraph $G' = (V', E')$ that enables solving the decision problem of whether there is a walk in $G$ matching an arbitrarily long query string $q$ in time linear in $|q|$ and independent of $|E|$ and $|V|$. We show $G$ is uniquely determined by $G'$ when for every $v_\ell \in V$, there is some distinct string $s_\ell$ on $\Omega$ such that $v_\ell$ is the origin of a closed walk in $G$ matching $s_\ell$, and no other walk in $G$ matches $s_\ell$ unless it starts and ends at $v_\ell$. We then exploit this invertibility condition to strategically alter any $G$ so its transform $G'$ enables retrieval of all $t$ terminal vertices of walks in the unaltered $G$ matching $q$ in $O(|q| + t \log |V|)$ time. We conclude by proposing two defining properties of a class of transforms that includes the Burrows-Wheeler transform and the transform presented here.
\end{abstract}

\section{Introduction}
\label{sec:intro}

Consider the decision problem of whether there is a walk in a finite edge-labeled digraph matching a query string of labels. Intuitively, the offline version of this problem is straightforwardly solved in time linear in the size of the string and independent of the size and order of the graph using an index that sorts the walks matching every possible query string so they can essentially be performed as if they were one walk. Many approaches \cite{siren2014indexing, novak2016graph, siren2017indexing, siren2020haplotype, garrison2018variation, prezza2021subpath, bowe2012succinct, gagie2017wheeler} further rely on the last-to-first (LF) mapping property exhibited by a class of invertible transforms \cite{ferragina2009compressing, mantaci2007extension} that includes the Burrows-Wheeler transform (BWT) \cite{burrows1994block}. Wheeler graphs \cite{gagie2017wheeler} provide a unifying formalism for these LF mapping-based strategies. A Wheeler graph admits a particular total order of its (unlabeled) vertices\footnote{in contradistinction to the order of a graph, which is the number of its vertices} such that the terminal vertices of all walks matching any query string comprise a single interval. LF mapping enables performing all walks in the Wheeler graph matching a given query string as if they were one walk from interval to interval of the totally ordered vertices.

But not every labeled digraph is a Wheeler graph. Indexing a non-Wheeler graph using the framework outlined above involves constructing an equivalent Wheeler graph of which subsets of vertices each represent a distinct vertex of the non-Wheeler graph. A query string is matched by a walk to a vertex of this original graph if and only if it is also matched by a walk to an associated vertex of the equivalent Wheeler graph. After confirming the presence of at least one match in the equivalent Wheeler graph, it is often important to perform a locate query, which retrieves the vertices of the original graph at which matching walks terminate. This requires additionally storing the vertex associations between the original graph and its equivalent Wheeler graph \cite{siren2014indexing, siren2017indexing, garrison2018variation}.

There are compressed indexes of Wheeler graphs supporting efficient locate queries \cite{gagie2017wheeler, gagie2021indexing, prezza2021locating}, but an equivalent Wheeler graph can be large and unwieldy regardless, especially when the non-Wheeler graph it represents has cycles. In practice, such cycles are typically indexed so only query strings up to some maximum size are supported \cite{siren2017indexing, garrison2018variation}, limiting the size and order of the equivalent Wheeler graph, but it is worth seeking alternative strategies that more readily and more elegantly accommodate matching query strings of arbitrary size and locating matches. The central issue is that LF mapping is a navigational instrument restricted to a line of vertices, mapping intervals into intervals. In this sense, a non-Wheeler graph is linearized by an equivalent Wheeler graph. Linearization may be cumbersome when the topology of the non-Wheeler graph deviates substantially from the topology of a Wheeler graph. Moreover, the terminal vertices of walks in a non-Wheeler graph matching a given query string are in general redundantly represented by the corresponding matching interval of totally ordered vertices of an equivalent Wheeler graph---that is, vertices on this interval may represent subsets of vertices of the non-Wheeler graph that are not disjoint. A locate query then returns a multiset of vertices of the non-Wheeler graph that must be deduplicated \cite{siren2014indexing, siren2017indexing, garrison2018variation}. The performance of such a query depends on the size of the multiset, which can significantly exceed the number of unique vertices it comprises. A full-featured navigational instrument would go beyond LF mapping and be able to map arbitrary subsets of vertices into arbitrary subsets of vertices, eliminating this performance bottleneck while also accommodating any finite labeled digraph.

The powerset construction \cite{rabin1959finite, hopcroft2001introduction} is just such an instrument, providing a transform of one labeled digraph into another labeled digraph of which each vertex represents a subset of vertices of the original graph. Moreover, this transformed graph can be used to solve the string matching decision problem with the same performance as an LF mapping-based framework. We show (1) under a particular condition, the original graph is uniquely determined by just the transformed graph, and (2) this invertibility condition can be exploited in a framework for efficiently locating query matches in \textit{any} finite labeled digraph, in analogy to how invertibility of the BWT for strings enables the FM-index \cite{ferragina2000opportunistic, ferragina2005indexing}, a widely used \cite{langmead2009ultrafast, li2009fast, langmead2012fast, li2013aligning} compressed suffix array.

Here is a brief summary of our framework: start with any finite digraph whose edges are labeled on some alphabet, and add edges to form a cycle that includes all vertices and matches a generalized de Bruijn sequence \cite{gabric2019generalized} on a different alphabet. As explained in Section~\ref{sec:discuss}, this operation is analogous to adding a sentinel (``the dollar sign'') to a string before obtaining its BWT. Think of the digraph as a nondeterministic finite automaton (NFA) whose every vertex is both an initial state and a final state and whose labeled edges encode the transition function. Use the powerset construction to obtain a deterministic finite automaton (DFA) from this NFA. Now think of the DFA as a digraph whose labeled edges encode the transition function and whose states are unlabeled vertices except for the empty state, which is excluded from the graph along with all states unreachable from the initial state. Refer to that initial state as the root. We call this transformed graph a nength of the original graph. There is at most one edge with a given label directed from any vertex of the nength. A walk in the original graph matches a query string if and only if a walk starting at the root of its nength also matches that string. When such a walk is present in the nength, traversals of the nength's edges labeled on the alphabet of the generalized de Bruijn sequence and starting at the terminal vertex of this walk rapidly and nonredundantly retrieve the terminal vertices of all walks in the original graph matching the query string.

Our main message is this: existing software \cite{siren2014indexing, siren2017indexing, garrison2018variation} for offline string matching in a labeled digraph linearizes that graph and subsequently indexes the result using BWT-based approaches familiar from text indexing. These approaches exploit invertibility of the BWT or a related transform via LF mapping to downsample vertex indices stored to support locate queries. However, when linearization is awkward and gives a massive equivalent Wheeler graph, downsampling is severely constrained and of diminished utility. In this event, it is possible to use a nength instead, which obviates the need for graph linearization and still exploits invertibility---achieved via a mechanism different from LF mapping---to reduce the index size while keeping locate queries efficient. Nength has the additional advantage that it naturally accommodates matching query strings of arbitrary size in any labeled digraph, whether or not it has cycles.

\begin{table}[ht]
\centering
\begingroup
\setlength{\tabcolsep}{6pt} 
\renewcommand{\arraystretch}{1.3} 
\begin{tabular}{rl}
\toprule
term & definition \\
\cmidrule(r){1-1}\cmidrule(lr){2-2}
\textbf{head} of a digraph's edge & vertex at which the edge is directed \\
\textbf{tail} of a digraph's edge & vertex from which the edge is directed \\
\textbf{order} of a graph & number of vertices of the graph; the word ``order'' is also \\ & used in this paper to refer to ordering objects, and the \\ & appropriate denotation should be clear from context \\ 
\textbf{size} of a graph & number of edges of the graph \\
\textbf{walk} in a digraph & sequence alternating between vertices and edges of the digraph \\ & such that each edge in the sequence is directed from the vertex \\ & immediately before it and at the vertex immediately after it\\
\textbf{closed walk} in a digraph & walk in the digraph that starts and ends at the same vertex \\
\textbf{path} in a digraph & walk in the digraph that repeats neither vertices nor edges \\
\textbf{necklace} & circular string of characters; if, e.g., $101$ is said to be a \\ & necklace, then the set of its two-character substrings is \\ & $\{10, 01, 11\}$, and $011$ and $110$ refer to the same necklace \\
\textbf{transform} & function, irrespective of its domain and codomain \\
\bottomrule
\end{tabular}
\endgroup
\vspace{3mm}
\caption{A glossary of terms used in this paper.}
\label{tab:terms}
\end{table}

Our presentation is organized as follows. Section~\ref{sec:transform} introduces the powerset construction using graph theoretic language and proves a general invertibility condition. Section~\ref{sec:loc} gives an algorithm for locating query matches with a nength, which relies on this invertibility condition via the generalized de Bruijn sequence construction sketched above, and describes a basic data structure for storing a nength. Section~\ref{sec:discuss} elaborates on the analogy between nength and the BWT and proposes two defining properties of a class of transforms that includes both. A glossary of terms required to understand this paper is provided in Table~\ref{tab:terms}. Other terms introduced here or invoked in more specific contexts than may be typical in the literature are italicized and defined on first use.

\section{Transform}
\label{sec:transform}

Let $G = (V, E)$ be a finite digraph where each vertex is unlabeled, each edge bears exactly one label on some alphabet $\Omega$, and any two edges with both the same head and the same tail have different labels. Let $G' = (V', E')$ be another digraph specified by $G$ according to the following conditions:
\begin{enumerate}
    \item each vertex $v'_i \in V'$ is unlabeled but associated with a distinct bit vector $\mathbf{b}'_i$ of size $|V|$ called the \textit{state} of $v'_i$ whose $\ell$th bit is $b'_{i\ell}$ and whose bits are never all zero,
    \item exactly one edge labeled $\omega_k \in \Omega$ extends from $v'_i$ to $v'_j$ for $v'_i, v'_j \in V'$ if and only if $\{ v_m \in V \, : \, b'_{jm} = 1\}$ is the set of heads of $\omega_k$-labeled edges of $G$ whose tails are among $\{ v_m \in V \, : \, b'_{im} = 1\}$, and
    \item $|V'|$ is as large as possible such that all vertices in $V'$ are reachable from a vertex designated as the root whose state has only nonzero bits.
\end{enumerate}
Condition 1 above implies $G'$ is finite because $G$ is finite, and there are $2^{|V|} - 1$ possible nonzero states. Condition 2 implies a vertex of $G'$ is the tail of no more than one edge labeled by a given character in $\Omega$. Condition 3 implies $G'$ is weakly connected.

A vertex $v'_i$ of $G'$ represents a set of vertices of $G$, and the state $\mathbf{b}'_i$ records these vertices. Note $G'$ can be thought of as the DFA obtained via the powerset construction \cite{rabin1959finite, hopcroft2001introduction} from, in general, an NFA. In the NFA, $\Omega$ is the set of input symbols, each state is both an initial state and a final state, each state corresponds to a distinct vertex of $G$, and the transition function is prescribed by the edges of $G$. In the DFA, the initial state corresponds to the root of $G'$. Further, every vertex of $G'$ corresponds to a distinct state of the DFA. States of the DFA unreachable from the initial state and transitions to the DFA's empty state are not represented in $G'$. See \cite{choi2013non, shao2014accelerating} for recent innovations in parallelization of the powerset construction.

$G'$ facilitates following all walks in $G$ matching some query string $q$ on $\Omega$ as if they were one walk: stand at the root of $G'$, start walking the sequence of edges whose labels match $q$, and either (1) it is not possible to reach step number $p \leq |q|$ because no walk in $G$ matches the size-$p$ prefix of $q$, or (2) the nonzero bits of the state of the vertex reached at step number $p$ correspond to the terminal vertices of all walks in $G$ matching the size-$p$ prefix of $q$. Since it essentially sorts all walks in $G$ and is obtained from the powerset construction, call $G'$ the \textit{powerset sort} of $G$. $G'$ permits solving the decision problem of whether there exist one or more walks in $G$ matching a query string $q$ in time linear in $|q|$, independent of $G$'s size $|E|$ and order $|V|$. If the states of vertices of the powerset sort $G'$ are stored beforehand, a positive determination is accompanied by the terminal vertices of matches in $G$. However, storing these states together with $G'$ is costly.

Call a vertex $v'_i \in V'$ for which $b'_{i\ell}$ is the only nonzero bit of $\mathbf{b}'_i$ the \textit{singleton} $v'^*_\ell$ of $G'$; that is, as an alternative notation, use an asterisk to denote singletons, and index them according to how corresponding vertices in $G$ are indexed. Now suppose for every $v_\ell \in V$, there is some distinct string $s_\ell$ on $\Omega$ such that $v_\ell$ is origin of some closed walk matching $s_\ell$, and no other walk in $G$ matches $s_\ell$ unless it starts and ends at $v_\ell$. Call $s_\ell$ an \textit{identifying string} of $v_\ell$; call a closed walk matching $s_\ell$ an \textit{identifying walk} of $v_\ell$. It is clear that if the state of each vertex of $G'$ is specified, $G$ is uniquely determined by $G'$: (1) for every vertex $v_\ell$ of $G$, there is a walk matching an identifying string of $v_\ell$ from $G'$'s root to the singleton $v'^*_\ell$; (2) for every $\omega_k \in \Omega$, if there are any $\omega_k$-labeled edges of $G$ whose tail is $v_\ell$, their heads are specified by the state of the head of the $\omega_k$-labeled edge extending from $v'^*_\ell$; and (3) this implies the head and tail of every edge of $G$ are known from $G'$ and the $\{\mathbf{b}'_w\}$. But a stronger statement can be made: $G'$ itself encodes the $\{\mathbf{b}'_w\}$ (up to permutation equivalence) via identifying walks, and it is an invertible transform of $G$ without requiring that the $\{\mathbf{b}'_w\}$ are recorded. We prove the following.
\begin{thm}\label{thm:inversion}
$G$ is uniquely determined by its powerset sort $G'$ when every vertex of $G$ has an identifying walk.
\end{thm}
\begin{proof}
$G'$ is given, but the state of each of its vertices is not. Write $\{v \rightarrow \}$ to refer to the set of strings matching walks starting at some vertex $v$ of some graph. Note by construction, $\{ v'^*_\ell \rightarrow \} = \{ v_\ell \rightarrow \}$ for $v'^*_\ell \in V'$ and $v_\ell \in V$; that is, the strings matching walks starting at a singleton of $G'$ capture precisely the set of possible matches to walks starting at its corresponding vertex in $G$. More generally,
\begin{equation}\label{eq:union}
    \{ v'_i \rightarrow \} = \bigcup_{w \in Y} \{v'^*_w \rightarrow \} \qquad \mbox {for } Y = \{ m \, : \, b'_{i m} = 1\}\,;
\end{equation}
that is, the strings matching walks starting at a given vertex $v'_i$ of $G'$ capture precisely the set of possible matches to walks starting at any vertex $v_\ell$ of $G$ for which $b'_{i \ell} = 1$. But by definition, for any singletons $v'^*_\ell, v'^*_p \in V'$ with $\ell \neq p$, $\{ v'^*_\ell \rightarrow \}$ contains an identifying string that is not in $\{ v'^*_p \rightarrow \}$. Together with \eqref{eq:union}, this says for any vertices $v_\ell \in V$ and $v'_i, v'_j \in V'$, $\{ v'_i \rightarrow \} \subseteq \{ v'_j \rightarrow \}$ if and only if $b'_{i \ell} = 1 \implies b'_{j \ell} = 1$. It follows that a given vertex $v'_i \in V'$ is a singleton if and only if for any vertex $v'_j \in V'$ with $j \neq i$, $\{ v'_j \rightarrow \} \not\subseteq \{ v'_i \rightarrow \}$. Further, $b'_{j \ell} = 1$ if and only if $\{ v'^*_\ell \rightarrow \} \subseteq \{ v'_j \rightarrow \}$ for $v'^*_\ell, v'_j \in V'$. This implies the states of all vertices of $G'$ can be determined up to permutation equivalence, and thus $G$ is uniquely determined by its powerset sort $G'$.
\end{proof}

\begin{figure*}[!t]
    \centering
    \includegraphics[width=\textwidth]{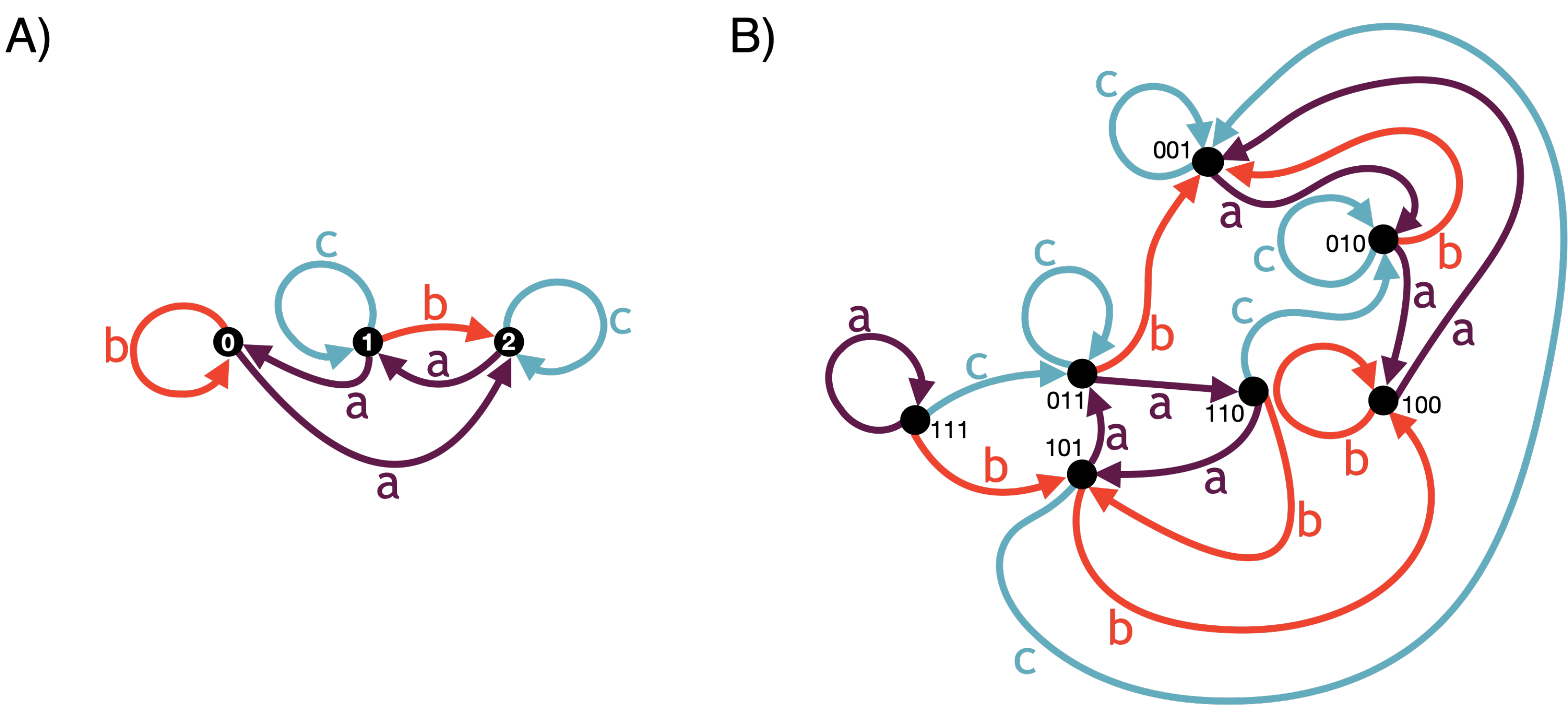}
    \caption{A) is an example digraph on the alphabet $\{a, b, c\}$ where each vertex has an identifying walk, and B) is its powerset sort. Edges with different labels have different colors. States are written next to associated vertices of B), and state bits are ordered correspondingly to vertex indices of A). A complete set of identifying strings for A) is $\{bb, cba, abc\}$. It is easily seen all walks in B) matching a given identifying string end at the same vertex, which is always a singleton.}
    \label{fig:exsort}
\end{figure*}

Let $\mathbf{A}_k$ be the adjacency matrix of $G$ specific to $\omega_k \in \Omega$; that is, its $(\ell, p)$th entry is $1$ when an $\omega_k$-labeled edge extends from $v_\ell$ to $v_p$ for $v_\ell, v_p \in V$ and is $0$ otherwise. Observe that $G'$ represents a system of matrix equations where multiplication is Boolean and a given edge labeled $\omega_k \in \Omega$ extending from $v'_i \in V'$ to $v'_j \in V'$ corresponds to the equation $\mathbf{A}_k \mathbf{b}'_i = \mathbf{b}'_j$. Perhaps surprisingly, Theorem \ref{thm:inversion} says this system has a unique solution up to permutation equivalence when it is constructed from a graph $G$ for which every vertex has an identifying walk, despite how none of the adjacency matrices or states is known in advance.

For a given identifying string $s_\ell$ of $v_\ell \in V$, any walk matching $s_\ell$ in $G'$ terminates at $v'^*_\ell$, and the origin of that walk has a state whose $\ell$th bit is nonzero because an identifying walk is closed. Call any set of identifying strings in which there is at least one identifying string per vertex of $G$ a \textit{complete set of identifying strings}. An example digraph is Figure~\ref{fig:exsort}A, and its powerset sort is Figure~\ref{fig:exsort}B. A complete set of identifying strings for Figure~\ref{fig:exsort}A is $\{bb, cba, abc\}$, and in Figure~\ref{fig:exsort}B, every walk matching any one of these identifying strings ends at the same singleton. The next section relies on Theorem~\ref{thm:inversion} to develop a framework for efficient location of matches to a query string in any finite labeled digraph.

\section{Location}
\label{sec:loc}

A de Bruijn sequence $B(r, n)$ of order $n$ on a size-$r$ alphabet is a necklace of size $r^n$ such that every possible size-$n$ string on the alphabet occurs exactly once as a substring. $B(r, n)$ is optimally short in the sense that a necklace of size $r^n$ has exactly as many substrings of size $n$ as there are possible strings of size $n$ on a size-$r$ alphabet. References \cite{gabric2019maximal, gabric2019generalized} introduce generalized de Bruijn sequences, a natural generalization of de Bruijn sequences to necklaces of arbitrary size. Let $x$ be some necklace, and let $\gamma_z(x)$ be the size of the set of size-$z$ substrings of $x$. A generalized de Bruijn sequence $B_G(r)$ on a size-$r$ alphabet for $r \leq |B_G(r)|$ is a necklace for which
$$
\gamma_z(B_G(r)) = \min(r^z, |B_G(r)|)\,.
$$
When $\gamma_z(B_G(r)) = |B_G(r)| = r^z$, $B_G(r)$ is a de Bruijn sequence of order $z$. Note $\lceil \log_{r} |B_G(r)| \rceil$ is the smallest value of $z$ such that $\gamma_z(B_G(r)) = |B_G(r)|$, and thus every size-$\lceil \log_{r} |B_G(r)| \rceil$ substring of $B_G(r)$ occurs exactly once as a substring of $B_G(r)$. References \cite{gabric2019maximal, gabric2019generalized} give a proof that there exists at least one generalized de Bruijn sequence $B_G(r)$ for any combination of $r \geq 2$ and $|B_G(r)| \geq 1$ and provide several examples of generalized de Bruijn sequences. Also refer to \cite{lempel1971m}, an antecedent with most of the elements of this proof.

Let $\widetilde{\Omega}$ be an alphabet of size at least $2$ such that $\widetilde{\Omega} \cap \Omega = \emptyset$. Call $\widetilde{\Omega}$ the \textit{sentinel alphabet}. Perform the following steps to alter any $G$ with at least two vertices\footnote{For $G$ with one vertex, $G'$ is invariably the same graph as $G$, making for a trivial case that need not be considered.} to form a new graph $\widetilde{G} = (V, \widetilde{E})$:
\begin{enumerate}
    \item Obtain some generalized de Bruijn sequence $c$ of size $|V|$ on $\widetilde{\Omega}$.
    \item Add edges to $G$ to form a cycle $G_C = (V, E_C)$ that includes every vertex and matches $c$.
\end{enumerate}
$\widetilde{G}$ is a labeled digraph on the alphabet $\Omega_U := \widetilde{\Omega} \cup \Omega$. Call the cycle subgraph $G_C$ of $\widetilde{G}$ the \textit{identifying cycle} of $\widetilde{G}$. Because $c$ is a generalized de Bruijn sequence, every vertex $v_\ell \in V$ is the origin of a walk in $G_C$ matching some size-$\lceil \log_{|\widetilde{\Omega}|} |V| \rceil$ string such that no other walk in $\widetilde{G}$ matches that string. This walk is part of a closed walk in the identifying cycle, and thus from Theorem \ref{thm:inversion}, $\widetilde{G}$'s powerset sort $\widetilde{G}' = (\widetilde{V}', \widetilde{E}')$ is invertible. Refer to the powerset sort of any digraph augmented with an identifying cycle as a \textit{nength} of that digraph. $\widetilde{G}'$ is a nength of $G$. Note that $G'$, the powerset sort of the original graph $G$, is a subgraph of the nength $\widetilde{G}'$: walks in $\widetilde{G}$ matching some query string $q$ on $\Omega$ end at vertices mirroring those in $G$ at which walks matching $q$ end.

Call a walk starting at any vertex $\widetilde{v}'_{i} \in \widetilde{V}'$ that is not a singleton a \textit{locating walk} of $\widetilde{v}'_{i}$ if it traverses only edges labeled on the sentinel alphabet $\widetilde{\Omega}$, ends at a singleton, and otherwise visits no singletons. Because it does not traverse any edges labeled on $\Omega$, a locating walk represents only walks in $\widetilde{G}$ confined to the identifying cycle; because it ends at a singleton, a locating walk represents exactly one walk in $\widetilde{G}$; because $G_C$ matches a generalized de Bruijn sequence, the size of a locating walk does not exceed $\lceil \log_{|\widetilde{\Omega}|} |V| \rceil$; because if any locating walk repeated a vertex, it would be possible to construct an arbitrarily long locating walk, a locating walk is always a path. The identifying cycle imposes a cyclic order on vertices of $\widetilde{G}$. Assign indices to vertices of $\widetilde{G}$ such that state bits respect this order. Then a locating walk of $\widetilde{v}'_{i}$ can be used to determine a nonzero bit of the state $\widetilde{\mathbf{b}}'_{i}$ of $\widetilde{v}'_{i}$, where a modular subtraction of the size of the locating walk from the index of its terminal singleton gives the index of the nonzero bit. This modular subtraction corresponds to a backwards walk in $G_C$ to recover the vertex starting the walk represented by the locating walk in $\widetilde{G}'$. The full state is recovered by following all locating walks of $\widetilde{v}'_{i}$. We prove the following.

\begin{prop}
Every $\widetilde{v}'_{i} \in \widetilde{V}'$ that is not a singleton has exactly as many locating walks in $\widetilde{G}'$ as there are nonzero bits of the state $\widetilde{\mathbf{b}}'_{i}$ of $\widetilde{v}'_{i}$, with each nonzero bit determined by a different locating walk.
\end{prop}
\begin{proof}
Suppose there were more locating walks than nonzero bits of $\widetilde{\mathbf{b}}'_{i}$. Then by the pigeonhole principle, there would be at least two distinct locating walks to singletons for which appropriate modular subtractions of steps from indices determined the same bit, some $\widetilde{b}'_{i\ell}$. But since both these walks represent walks in the identifying cycle $G_C$ starting at $v_\ell \in V$, (1) if they had the same number of the steps, they would necessarily correspond to the same walk in $G_C$, a contradiction, and (2) if they had different numbers of steps, the longer walk would reach a singleton before its end, a contradiction.
\end{proof}

So $\widetilde{G}'$ offers a straightforward way to obtain the vertices of $G$ matching any size-$p$ prefix of a query string $q$ on $\Omega$: stand at the root of $\widetilde{G}'$, start walking the sequence of edges whose labels match $q$, and if step $p$ is reached at some vertex, follow locating walks and perform appropriate modular subtractions to obtain the state of that vertex, whose nonzero bits correspond to the terminal vertices of walks in $G$ matching the size-$p$ prefix of $q$.

It is not necessary to store all of $\widetilde{G}'$ to enable these locate queries. Call a vertex of $\widetilde{G}'$ a \textit{spanner} if it is not a singleton and either is reachable from the root by following only edges labeled on $\Omega$ or is the root itself. Call a vertex of $\widetilde{G}'$ a \textit{locator} if it is not a singleton and is not reachable from the root by following only edges labeled on $\Omega$, but is reachable on some locating walk of a spanner. Store only singletons, spanners, locators, edges labeled on $\Omega$ whose tails are singletons, edges labeled on $\Omega_U$ whose tails are spanners, and edges labeled on $\widetilde{\Omega}$ whose tails are locators. Any other components of $\widetilde{G}'$ are not visited or traversed during string matching or on locating walks. Further, note it is enough to know only that an edge is labeled on the sentinel alphabet $\widetilde{\Omega}$ rather than $\Omega$ to follow locating walks; the particular label of an edge on $\widetilde{\Omega}$ need not be recorded.

Let $\mathbf{M}$ be a matrix with $|\Omega_U|$ columns where (1) each row corresponds to a different vertex of $\widetilde{G}'$, (2) each column corresponds to a different character in $\Omega_U$, (3) an entry is the null pointer if and only if there is no edge of $\widetilde{G}'$ whose tail corresponds to the entry's row and whose label corresponds to the entry's column, and (4) an entry is a pointer to some row if and only if that row corresponds to the head of an edge of $\widetilde{G}'$ whose tail corresponds to the entry's row and whose label corresponds to the entry's column. $\mathbf{M}$ can be used to perform a walk in $\widetilde{G}'$ by following pointers from row to row. Arrange the row order of $\mathbf{M}$ so its first $|V|$ rows correspond to singletons, and ensure these vertices are in an order prescribed by $G_C$. This implicitly stores their indices---that is, when a walk in $\widetilde{G}'$ using $\mathbf{M}$ ends at some ($0$-indexed) row $\ell < |V|$, that row corresponds to a singleton whose index according to $G_C$ is $\ell$. Arrange that the root of $\widetilde{G}'$ corresponds to the row of $\mathbf{M}$ right after the first $|V|$ rows so pattern matching always starts there. Also arrange that all rows corresponding to spanners precede all rows corresponding to locators, and all columns corresponding to characters in $\widetilde{\Omega}$ precede all columns corresponding to characters in $\Omega$. Further, since the particular labels of edges labeled on $\widetilde{\Omega}$ are inconsequential, reorder the first $\widetilde{\Omega}$ entries of each row of $\mathbf{M}$ so all nonnull pointers precede all null pointers.

\begin{figure}
\begin{algorithm}[H]
 \caption{State determination}
 \begin{algorithmic}[1]
 \renewcommand{\algorithmicrequire}{\textbf{Input:}}
 \renewcommand{\algorithmicensure}{\textbf{Output:}}
 \REQUIRE $\mathbf{M}$, index $i$ of row of $\mathbf{M}$ corresponding to vertex $\widetilde{v}'_{i}$ of $\widetilde{G}'$ whose state is desired, $|V|$, $|\widetilde{\Omega}|$
 \ENSURE size-$|V|$ state $\widetilde{\mathbf{b}}'_{i}$ of $\widetilde{v}'_{i}$
 \\ \textit{Initialization} : state $\widetilde{\mathbf{b}}'_{i} \leftarrow \mathbf{0}$, stack $S \leftarrow \{\}$
  \IF {$i < |V|$}
  \STATE $\widetilde{b}'_{ii} \leftarrow 1$
  \RETURN $\widetilde{\mathbf{b}}'_{i}$
  \ENDIF
  \FOR {$k := 0$ \TO $|\widetilde{\Omega}| - 1$}
  \IF {$M_{ik}$ is null}
  \STATE \textbf{break}
  \ENDIF
  \STATE push the tuple $(M_{ik}, 1)$ onto $S$
  \ENDFOR
  \WHILE{$S$ is not empty}
  \STATE pop some $(m, p)$ off $S$
  \IF {$m < |V|$}
  \STATE $\widetilde{b}'_{i(m - p)} \leftarrow 1$
  \ELSE
  \FOR {$k := 0$ \TO $|\widetilde{\Omega}| - 1$}
  \IF {$M_{mk}$ is null}
  \STATE \textbf{break}
  \ENDIF
  \STATE push the tuple $(M_{mk}, p + 1)$ onto $S$
  \ENDFOR
  \ENDIF
  \ENDWHILE
  \RETURN $\widetilde{\mathbf{b}}'_{i}$
 \end{algorithmic}
\end{algorithm}
\caption{A depth-first approach to determining the state of a vertex of some nength $\widetilde{G}'$. A given entry of $\mathbf{M}$ that is a nonnull pointer is taken to be the index of the row of $\mathbf{M}$ pointed. The first row of $\mathbf{M}$ corresponds to the index $0$. Bit indices of bit vectors respect congruence modulo $|V|$.}
\label{alg:stateres}
\end{figure}

The entries of $\mathbf{M}$ necessary for pattern matching can now be stored as an array $\mathbf{m}$ in row-major order, where there is: (1) a single block of rows corresponding to a singletons, with each row taking $|\Omega|$ elements; (2) a single block of rows corresponding to spanners, with each row taking $|\Omega_U|$ elements; and (3) a single block of rows corresponding to locators, with each row taking $|\widetilde{\Omega}|$ elements. Straightforward pointer arithmetic then gives the location in memory of any entry of $\mathbf{M}$ required for string matching or state determination. Our procedure for state determination is formalized in Figure~\ref{alg:stateres}. The logic for retrieving specific entries of $\mathbf{M}$ using $\mathbf{m}$ is excluded there.

Because of how a subgraph of $\widetilde{G}'$ is $G'$, $\widetilde{G}'$ has $|V'| - |V|$ spanners. The number of singletons of $\widetilde{G}'$ is the number $|V|$ of vertices of $\widetilde{G}$. Suppose $\widetilde{G}'$ has $|\widetilde{V}'_L|$ locators. Then $\mathbf{m}$ takes up
$$
\left(|\Omega_U| \left(|V'| - |V|\right) + |\Omega| |V| + |\widetilde{\Omega}||\widetilde{V}'_L|\right) \lceil \log_2 (|V'| + |\widetilde{V}'_L| + 1) \rceil
$$
bits.

\begin{figure*}[!t]
    \centering
    \includegraphics[width=\textwidth]{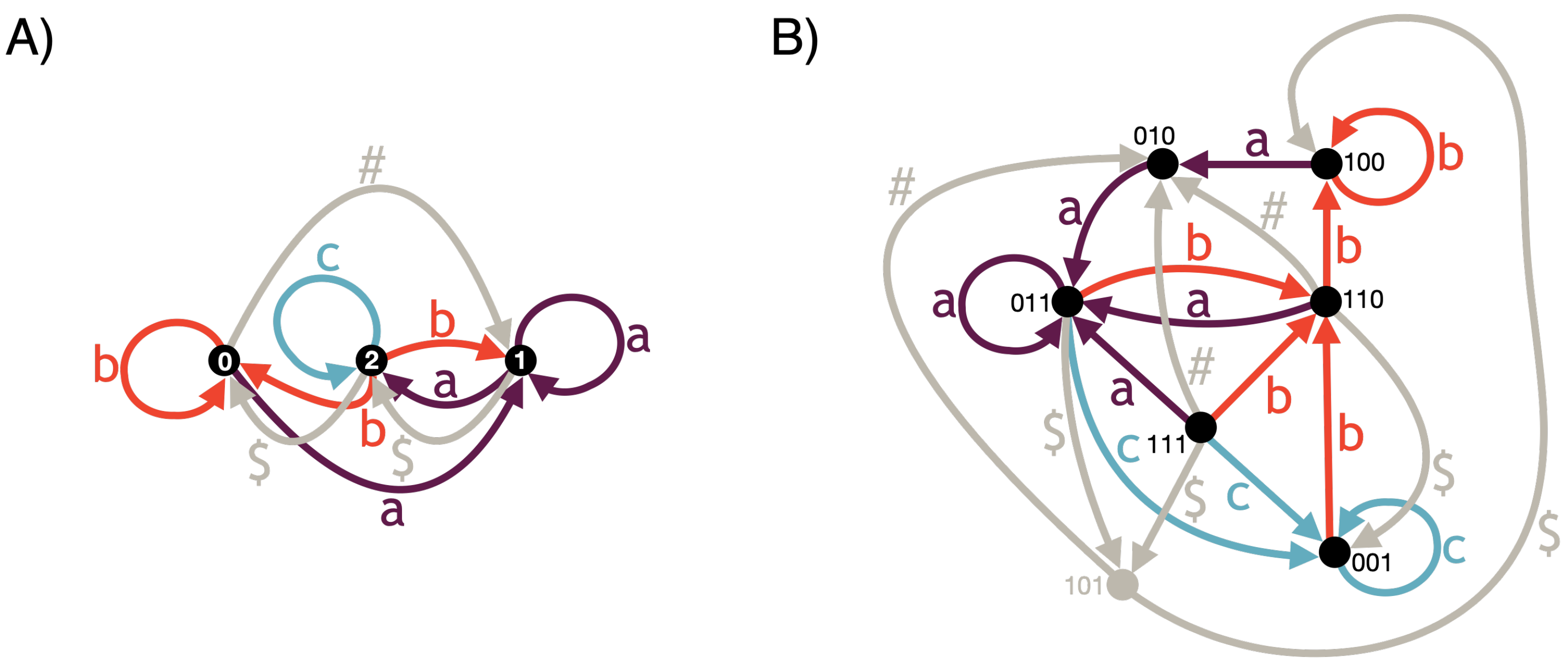}
    \caption{A) is a digraph with an identifying cycle on the sentinel alphabet $\{\$,\#\}$ matching the generalized de Bruijn sequence $\$\$\#$. B) is the corresponding nength, excluding edges labeled on $\widetilde{\Omega}$ whose tails are singletons and edges labeled on $\Omega$ whose tails are the sole locator (i.e., the vertex with state $101$). States are written next to associated vertices of B), and state bits are ordered correspondingly to vertex indices of A). Note the identifying cycle imposes an order on the vertices respected by their indices. Edges comprising the identifying cycle of A) and the graph components they contribute to B) are in gray. Otherwise, edges are assigned colors according to their labels.}
    \label{fig:exnength}
\end{figure*}

Figure~\ref{fig:exnength}A is a graph with an identifying cycle on the sentinel alphabet $\{\$, \#\}$. Without the identifying cycle, vertex $0$ has no identifying walk. Figure~\ref{fig:exnength}B is the corresponding nength, excluding edges labeled on $\widetilde{\Omega}$ whose tails are singletons and edges labeled on $\Omega$ whose tails are the sole locator (i.e., the vertex with state $101$).

Assume $|\widetilde{\Omega}| < |V|$ and that entries of $\mathbf{M}$ can be accessed in constant time. In the worst case, our algorithm for state determination runs in $O(t \log_{|\widetilde{\Omega}|} |V|)$ time, where $t$ is the number of nonzero bits of the state, since a locating walk can take up to $\lceil \log_{|\widetilde{\Omega}|} |V| \rceil$ steps. Identifying the $t$ terminal vertices of walks in $G$ matching a query string $q$ by following pointers in $\mathbf{M}$ thus takes $O(|q| + t \log_{|\widetilde{\Omega}|} |V|)$ time. The size of the sentinel alphabet $\widetilde{\Omega}$ can be as small as $2$, with larger alphabet sizes improving the performance of state determination while increasing the number of columns of $\mathbf{M}$.

By contrast, naively storing all states of the powerset sort $G'$ of $G$ in a $|V'| \times |V|$ binary matrix for their immediate retrieval gives $O(|q|)$ performance. But storing $|V'||V|$ state bits may be forbidding for large $G$, and the storage overhead of locating walks in the array $\mathbf{m}$ may be comparatively small for small $|\widetilde{\Omega}|$; for example, a single locator may be the head of many edges labeled on $\widetilde{\Omega}$, achieving compression by simultaneously representing configurations of state bits that are the same across the edges' tails. 

It is not absolutely necessary to ensure $G_C$ matches a generalized de Bruijn sequence to obtain the $O(t \log_{|\widetilde{\Omega}|} |V|)$ performance guarantee for state determination. References \cite{gabric2019maximal, gabric2019generalized} provide an alternative characterization of a generalized de Bruijn sequence $B_G(r)$ as a necklace on a size-$r$ alphabet that satisfies
\begin{enumerate}
    \item $\gamma_{d}(B_G(r)) = r^d$
    \item $\gamma_{d+1}(B_G(r)) = |B_G(r)|$
\end{enumerate}
with $d := \lfloor \log_{r} |B_G(r)| \rfloor$. A necklace satisfying condition 2 above but not necessarily condition 1 also gives the performance guarantee. For some intuition about the difference, consider the following example borrowed from \cite{gabric2019maximal, gabric2019generalized}: on the binary alphabet $\{0, 1\}$, $00001011101$ is a generalized de Bruijn sequence, but $10011110000$ satisfies only condition 2. While every size-$4$ substring is distinct in each sequence, the former has all eight possible size-$3$ substrings, while the latter has only seven and is missing $101$. A necklace satisfying condition 2 but not necessarily condition 1 is called an $m$-ary closed sequence in \cite{lempel1971m}, and construction algorithms were developed decades ago \cite{hemmati1978algebraic, etzion1986algorithm}. However, it is desirable that $G_C$ matches a necklace that also satisfies condition 1. Maximizing complexity in this way can in general reduce the average number of steps of a locating walk: fewer instances of particular short kmers in the necklace make for fewer steps along the identifying cycle to distinguish vertices. So it is worth exploring how to construct generalized de Bruijn sequences efficiently.

Our basic data structure for storing $\mathbf{M}$ can be refined to reduce its size. A degree of freedom we do not explore thoroughly here is that rows corresponding to locators and spanners can be reordered so $\mathbf{M}$ has more structured sparsity. For example, ordering rows of $\mathbf{M}$ to cluster them according to which of their columns contains null pointers permits eliminating null pointers from the array $\mathbf{m}$ if the indices of nonnull columns on intervals of rows are recorded in an auxiliary data structure. Rows can also be ordered lexicographically by treating them like they are strings on an ordered alphabet of pointers; intervals of rows with the same prefix can then be compressed. It may also be possible to reduce the total number of rows of $\mathbf{M}$ by arranging that the identifying cycle largely follows existing paths in $G$; when $G$ is suitably sparse, this could give rise to $\widetilde{G}'$ with a preponderance of pairs of edges sharing the same head and tail, thereby economizing the number of its vertices. The designs of identifying cycles and of compressed representations of $\mathbf{M}$ are thus potentially fruitful areas for further research.

\section{Discussion}
\label{sec:discuss}

Consider the case where $G$ is a cycle graph whose every vertex has an identifying walk, and sort the $|E|$ identifying strings in lexicographic order, writing the result as an $|E| \times |E|$ matrix $\mathbf{B}$. The $i$th instance of a given character $\omega_k \in \Omega$ in the first (F) column of $\mathbf{B}$ corresponds to the same edge as the $i$th instance of $\omega_k$ in the last (L) column of $\mathbf{B}$. This LF mapping property means that $G$ is implicitly encoded in $\mathbf{B}$'s last column, which is the BWT of $G$. To recover $G$ from its BWT, note first the F column is exactly the characters of the L column in sorted order---that is, F is composed of successive blocks of characters from $\Omega$, with one block per character. Write the F and L columns next to each other. Now:
\begin{enumerate}
\item start at some arbitrary row,
\item apply LF mapping to the character in the F column to move to the row whose character in the L column corresponds to the same edge, and
\item repeat step 2 until the starting row of step 1 is reached.
\end{enumerate}
The sequence of characters in the F column encountered on following these instructions completely recapitulates the cycle comprising $G$. If a given query string $q$ on $\Omega$ matches several walks in $G$, all these walks correspond to a single interval of rows of $\mathbf{B}$. LF mapping can be applied to obtain this interval in time linear in $|q|$, independent of $|E|$ and $|V|$, with an appropriately designed rank data structure over the BWT \cite{ferragina2000opportunistic}.

LF mapping is generally encountered as a byproduct of an ordering procedure. A recent generalization \cite{cotumaccio2021indexing} of the BWT that applies to an arbitrary labeled digraph $G$ obtains a partial co-lexicographic order of its vertices. In this framework, walks in $G$ matching a query string always terminate at some convex subset of the ordered vertices. In the worst case, search performance using a proposed extension to the FM-index based on this transform is $O(|q||E|\log |V|)$. The result is consistent with a recently obtained conditional lower bound for string matching in labeled digraphs: unless the strong exponential time hypothesis is false, no index constructed in time polynomial in $|E|$ can deliver a search performance of $O(|q|^\delta |E|^\beta )$ with either $\delta < 1$ or $\beta < 1$ \cite{equi2020conditional, equi2021graphs}.

A nength can have up to $2^{|V|} - 1$ vertices, so the asymptotic scalings of both its construction time and index size include an exponential factor whose argument is $|V|$. However, we expect the situation is not so grim for many classes of graphs. Indeed, \cite{cotumaccio2021indexing} establishes an upper bound of $2^p (|V| - p + 1) - 1$ on the number of states of the DFA obtained by applying the powerset construction to an NFA with $|V|$ states, where $p$ is the width of a partial co-lexicographic order of the NFA's states. Reference \cite{cotumaccio2021indexing} further notes the parameter $p$ serves as a complexity measure for graphs, where Wheeler graphs have $p = 1$.

Various analogs to the BWT respecting some of its features while discarding others are possible. In the analog described in \cite{cotumaccio2021indexing}, invertibility is achieved in the general case by explicitly recording a strategic abbreviation of submatrices of the graph's adjacency matrix that exploits the partial co-lexicographic order of vertices. As the complexity measure $p$ increases, this representation collapses to exactly the adjacency matrix. So the representation is guaranteed to be invertible because at worst, it is a literal encoding of the original graph. In the analog to the FM-index built on this representation, compression is achieved at the expense of search performance, both of which degrade as $p$ increases.

Our perspective is that the powerset construction itself provides a BWT-like transform. A nength sorts possible matches without ordering them. While it no longer has a semblance of what makes the BWT navigable---LF mapping---what makes the BWT invertible is preserved: every vertex has an identifying walk. (Note, for example, if multiple closed walks in a cycle graph matched the same string, the BWT matrix $\mathbf{B}$ would be a sequence of blocks of identical rows. LF mapping would then obtain multiple cycles rather than a single cycle, and invertibility would be lost.) Just as the BWT is an invertible transform of a cycle graph into a string with a beginning and an end, a result of ordering, a nength is an invertible transform of a labeled graph into a different graph for which each vertex is the tail of at most one edge labeled by a given character, a result of sorting.

For both the BWT and nength, properties linked to invertibility can be exploited to rapidly locate matched patterns. An arbitrary finite string on $\Omega$ can be extended by an extra sentinel character that is not in $\Omega$. The ends of this string can then be joined to form an aperiodic necklace. The BWT of this necklace is invertible because each character has a distinct distance from the sentinel. The identifying cycle labeled on the sentinel alphabet performs the same function for a nength as a sentinel does for a BWT; there is not necessarily a natural distance between two given vertices of an arbitrary graph, but adding an identifying cycle vests the graph with a distance function on its vertices. Ensuring this cycle matches a generalized de Bruijn sequence gives a performance guarantee for state determination via nength navigation.

The paper introducing Wheeler graphs \cite{gagie2017wheeler} articulates two main features of the original BWT: (1) it is invertible, and (2) it ``helps'' compression. The paper also notes some variants of the BWT in the literature do not have these features. Since there are indeed so many such variants, it is worth considering how to define properties of a potentially broad class of transforms that includes both the BWT and nength. We believe references to compression should be avoided. A labeled digraph can be thought of as a potentially compressed representation of many strings, apart from any transform. Moreover, the BWT does not itself do any compressing, and that it tends to help in approaches to lossless compression of text is, of course, an artifact of the distribution of data encountered in practical settings; there exists some distribution of data for which it would typically ``hurt'' compression. Rather, we believe at its core, the BWT is a tool for maximally efficient string matching. We also believe invertibility alone is not one of its defining properties. How the BWT achieves invertibility matters.

Given these considerations, we propose defining a search transform as follows. Let $X$ be a configuration of unlabeled objects together with directed relationships, where each relationship connects a subset of objects and has a set of labels, potentially on multiple alphabets. A \textit{search transform} is any transform of $X$ into a different configuration $X'$ of objects and relationships such that
\begin{enumerate}
    \item $X'$ enables an index that answers whether a structured query pattern of relationships is present in $X$ in time independent of the numbers of objects and relationships $X$ contains, and
    \item $X$ is uniquely determined by $X'$ precisely because for every object in $X$, there is some nonempty query pattern matched only at that object.
\end{enumerate}
Above, we draw a distinction between a search transform and an index enabled by a search transform. A total order of the vertices of a Wheeler graph together with auxiliary data supporting LF mapping-based navigation is a search transform because an FM-index built on it solves the string matching decision problem in time independent of the size and order of the graph \cite{gagie2017wheeler}. This is despite how using the more compact $r$-index \cite{gagie2018optimal, gagie2020fully, navarro2019universal, kempa2019optimal, kuhnle2020efficient, bannai2020refining} in place of the FM-index solves the string matching decision problem in time polylogarithmic in the graph's size \cite{gagie2021indexing}. Both nength and the BWT are also search transforms. However, the compound transform that linearizes a non-Wheeler graph and subsequently orders the vertices of an equivalent Wheeler graph is excluded from our definition because it is invertible in part via the map between the non-Wheeler graph and the equivalent Wheeler graph. Note our definition leaves room for possible transforms that facilitate matching of patterns more involved than strings. We leave exploration of these possibilities for future work.

\section*{Acknowledgements}
We thank Rachel Ward, Ben Langmead, Chris Wilks, and the anonymous reviewers for the 32nd Annual Symposium on Combinatorial Pattern Matching, all of whose feedback improved this paper considerably.

\bibliographystyle{unsrt}
\bibliography{references}

\end{document}